\documentclass[a4paper,fleqn,usenatbib]{mnras}

\usepackage{newtxtext,newtxmath}

\usepackage[T1]{fontenc}
\usepackage{ae,aecompl}

\usepackage{graphicx}	
\usepackage{amsmath}	
\usepackage{amssymb}	

\title[Molecular Outflows of NGC~6240 with ALMA]{Imaging the Molecular Outflows of the Prototypical ULIRG NGC~6240 with ALMA}

\author[Toshiki Saito et al.]{
T. Saito,$^{1,2}$\thanks{E-mail: saito@mpia-hd.mpg.de}
D. Iono,$^{1,3}$
J. Ueda,$^{4}$
D. Espada,$^{1,3}$
K. Sliwa,$^{2}$
K. Nakanishi,$^{1,3}$
N. Lu,$^{5,6}$
\newauthor
C. K. Xu,$^{5,6}$
T. Michiyama,$^{1,3}$
H. Kaneko,$^{1,7}$
T. Yamashita,$^{8}$
M. Ando,$^{1,3}$
M. S. Yun,$^{9}$
\newauthor
K. Motohara,$^{10}$
and R. Kawabe$^{1,3}$
\\
$^{1}$National Astronomical Observatory of Japan, 2-21-1 Osawa, Mitaka, Tokyo, 181-8588, Japan\\ 
$^{2}$Max-Planck-Institut f\"ur Astronomie, K\"onigstuhl 17, 69117 Heidelberg, Germany\\
$^{3}$The Graduate University for Advanced Studies (SOKENDAI), 2-21-1 Osawa, Mitaka, Tokyo 181-8588, Japan\\
$^{4}$Harvard-Smithsonian Center for Astrophysics, 60 Garden Street, Cambridge, MA 02138, USA\\
$^{5}$National Astronomical Observatories of China, Chinese Academy of Sciences, Beijing 100012, China\\
$^{6}$China-Chile Joint Center for Astronomy, Chinese Academy of Sciences, Camino El Observatorio, 1515 Las Condes, Santiago, Chile\\
$^{7}$Nobeyama Radio Observatory, National Astronomical Observatory of Japan, Minamimaki, Minamisaku, Nagano 384-1305, Japan\\
$^{8}$Research Center for Space and Cosmic Evolution, Ehime University, 2-5 Bunkyo-cho, Matsuyama, Ehime 790-8577, Japan\\
$^{9}$Department of Astronomy, University of Massachusetts, Amherst, MA 01003, USA\\
$^{10}$Institute of Astronomy, The University of Tokyo, 2-21-1 Osawa, Mitaka, Tokyo 181-0015, Japan
}

\date{Accepted XXX. Received YYY; in original form ZZZ}

\pubyear{2017}

\begin{document}
\label{firstpage}
\pagerange{\pageref{firstpage}--\pageref{lastpage}}
\maketitle

\begin{abstract}
We present 0\farcs97 $\times$ 0\farcs53 (470~pc $\times$ 250~pc) resolution CO~($J$ = 2--1) observations toward the nearby luminous merging galaxy NGC~6240 with the Atacama Large Millimeter/submillimeter Array.
We confirmed a strong CO concentration within the central 700~pc, which peaks between the double nuclei, surrounded by extended CO features along the optical dust lanes ($\sim$11~kpc).
We found that the CO emission around the central a few kpc has extremely broad velocity wings with full width at zero intensity $\sim$ 2000 km s$^{-1}$, suggesting a possible signature of molecular outflow(s).
In order to extract and visualize the high-velocity components in NGC~6240, we performed a multiple Gaussian fit to the CO datacube.
The distribution of the broad CO components show four extremely large linewidth regions ($\sim$1000 km s$^{-1}$) located 1-2~kpc away from both nuclei.
Spatial coincidence of the large linewidth regions with H$\alpha$, near-IR H$_2$, and X-ray suggests that the broad CO~(2--1) components are associated with nuclear outflows launched from the double nuclei.

\end{abstract}

\begin{keywords}
galaxies: individual (NGC~6240) --- galaxies: interactions --- galaxies: evolution --- galaxies: active
\end{keywords}



\section{Introduction} \label{S_1}
Gas-rich galaxy mergers play a major role in the formation and evolution of galaxies by triggering intense star formation and changing their morphology as suggested by numerical simulations \citep[e.g.,][]{Barnes91}.
During the process of a galaxy merger, radial streaming can feed gas to the central supermassive black hole (SMBH) \citep[e.g.,][]{Hopkins08}, possibly reaching at the quasar-phase after dispersing the obscuring gas and dust \citep{Urrutia08}.
Powerful winds or outflows, driven by the active galactic nucleus (AGN) or the surrounding starbursts ((ultra-)luminous infrared galaxy, (U)LIRG), are predicted to suppress gas feeding to the central SMBH and its host galaxy's spheroidal component in order to explain the empirical correlation between the stellar velocity dispersion of a galaxy bulge and the mass of the central SMBH \citep{Costa14}.

Observational studies of such galactic winds/outflows in (U)LIRGs have been frequently done by spectral analyses of hot molecular gas, atomic gas, and ionized gas emission/absorption lines \citep[e.g.,][]{Bellocchi13,Veilleux13}.
Those tracers are bright enough to probe generally faint, broad outflowing gas profiles, although they are not likely to be major constituents of galactic outflows.
Cold molecular gas is likely to dominate the outflow in terms of mass, and thus it is the key to understanding fundamental feedback effects of galactic outflows on the surrounding interstellar medium (ISM) \citep{Feruglio10,Cicone14,Fiore17}.

NGC~6240 is a well-studied close galaxy pair in the local Universe (z = 0.024480, 1\farcs0 = 480~pc), which is often regarded as a prototype of ULIRGs, even though its infrared (IR) luminosity is slightly lower than 10$^{12}$~$L_{\odot}$ \citep[$L_{\rm IR}$ = 10$^{11.93}$~$L_{\odot}$;][]{Armus09}.
This system has a double nucleus detected in a variety of wavelengths from X-ray to radio \citep[e.g.,][]{Komossa03,Hagiwara11,Stierwalt14,Ilha16}, indicating the presence of two separated AGNs.
The cold molecular ISM in NGC~6240 has been observed through many rotational transitions of CO, HCN, and HCO$^+$, showing a strong gas concentration between the nuclei \citep[e.g.,][]{Nakanishi05,Iono07,Papadopoulos14,Scoville15,Tunnard15,Sliwa17}.
Although the properties shown above are indeed suitable for representing nearby ULIRGs, past observations have revealed a different side of NGC~6240, which makes it rather unique.
\citet{Lu15} reported that NGC~6240 shows an order of magnitude higher $L_{\rm CO(7-6)}$/$L_{\rm IR}$ ratio than other local (U)LIRGs.
Such an extreme CO line-to-continuum ratio can be explained by a simple C-shock model \citep{Meijerink13}.
Furthermore, deep and wide Subaru observations revealed a vastly extended, filamentary H$\alpha$ nebulae \citep[$\sim$90~kpc;][]{Yoshida16} coinciding with soft X-ray emission.
Contrary to those extended features, the H$\alpha$ and X-ray observations have also detected a compact ``butterfly" nebula ($\sim$4.5~kpc) surrounding the nuclei showing the signatures of ionized gas outflows \citep{Komossa03}.
Recent interferometric observations have revealed the presence of a broad molecular line profile ($\leq$ 1400~km~s$^{-1}$) toward the center \citep{Tacconi99,Ohyama03,Iono07,Feruglio13a,Feruglio13b}.
However, the spatial and velocity structures of the central kpc of NGC~6240 are still unclear mainly because of the limited sensitivity, angular resolution, and/or bandwidth of previous observations.
Also, the complex velocity structures due to the ongoing violent merging event prevent us from modeling and revealing the underlying gaseous structures.

In this letter, we present Atacama Large Millimeter/submillimeter Array (ALMA) Band~6 observations of NGC~6240 in the CO~(2--1) line emission and its underlying continuum with sub-arcsecond resolution.
The goal of this letter is to reveal the complete spatial and velocity distribution of the molecular outflow, which has been previously only partly mapped, in NGC~6240.
We assumed $H_0$ = 70~km~s$^{-1}$~Mpc$^{-1}$, $\Omega_{\rm M}$ = 0.3, $\Omega_{\rm \Lambda}$ = 0.7 throughout this letter.

\section{ALMA Band~6 Observation}
NGC~6240 was observed on 2016 June 26 for the ALMA Cycle~3 program ID = 2015.1.00003.S using forty-two 12~m antennas.
The Band~6 receiver was tuned to cover CO~(2--1) ($\nu_{\rm obs}$ = 225.02928~GHz).
The field of view, on-source time, and baseline lengths are $\sim$28\arcsec, $\sim$4 minutes, and 15.1--783.5~m, respectively.
One of the spectral windows (spw) containing CO has a resolution of 1.953~MHz ($\sim$2.6~km~s$^{-1}$) with a bandwidth of 1.875~GHz.
The single sideband system temperature after flagging is 50--100~K with two peaks of $\sim$190~K at atmospheric absorptions.
Precipitable water vapour during the observations toward NGC~6240 and all calibrators are 0.68--0.82~mm.
Titan, J1550+0527, and J1651+0129 were observed as the amplitude, bandpass, and phase calibrators, respectively.
We used the delivered calibrated $uv$ data.
The flux of J1550+0527 at each spw is in agreement with the ALMA Calibrator Source Catalogue\footnote{https://almascience.nrao.edu/sc/} (a flux measurement obtained on the same date).
Thus, we regard the accuracy of the absolute flux calibration as 5\% throughout this paper.

The data reduction was carried out using {\tt CASA} version 4.5.3 \citep{McMullin07}.
All maps are reconstructed with natural weighting.
We used the {\tt CASA} task {\tt clean} in multi-scale mode to make use of the multiscale CLEAN deconvolution algorithm \citep[MS-CLEAN;][]{Cornwell08}.
MS-CLEAN can reduce extended low surface brightness structures in the residual image, and thus recover extended emission significantly larger than the synthesized beam size.
For the deconvolution process using {\tt clean}, we performed an iterative auto-masking procedure as described in the CASA guide\footnote{https://casaguides.nrao.edu/index.php/M100\_Band3\_Combine\_\\4.3\#Image\_Using\_An\_Automasking\_Technique}.
In order to improve the image fidelity, we carried out a two-rounded phase self-calibration after the normal ALMA calibration process.
We chose the bright compact source corresponding to the peak of the CO spectrum ($\sim$7050~km~s$^{-1}$) for the model used in the self-calibration.
Continuum emission was subtracted in the $uv$-plane by fitting the line-free channels in both lower and upper sidebands with a first order polynomial function, and then we made a CO~(2--1) data cube with 5~km~s$^{-1}$ resolution.
The synthesized beam size and sensitivity are 0\farcs97 $\times$ 0\farcs53 (PA = 64\fdg5) and 2.5 mJy beam$^{-1}$ (at 5~km~s$^{-1}$ bin), respectively.

\begin{figure*}
	\begin{center}
	\includegraphics[width=1.8\columnwidth]{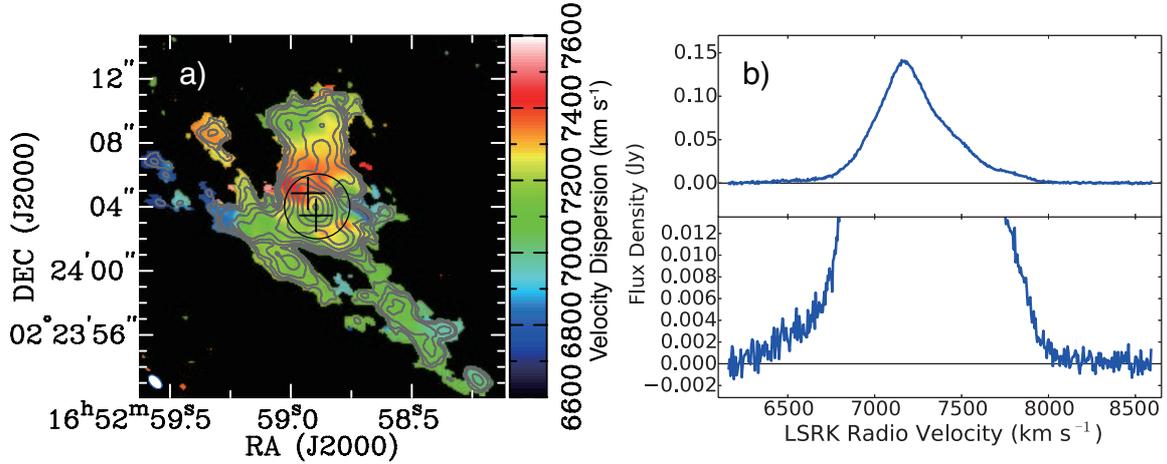}
   \end{center}
    \caption{(a) CO~(2--1) velocity field (moment~1) image of NGC~6240.  The velocity field in color scale ranges from 6600 to 7600 km s$^{-1}$.  Contours show the CO~(2--1) integrated intensity (moment~0) map.  The contours are 208 $\times$ (0.125, 0.25, 0.5, 1.0, 2.0, 4.0, 8.0, 16.0, 32.0 64.0, and 96.0) Jy beam$^{-1}$ km s$^{-1}$.  The crosses indicate the peak positions of the Band~6 continuum emission.  (b) CO~(2--1) spectrum toward the central 4\arcsec aperture, which is shown in Figure~\ref{fig_1}a, with a zoomed panel to emphasize the wings.}
    \label{fig_1}
\end{figure*}

\begin{figure}
	\begin{center}
	\includegraphics[width=0.9\columnwidth]{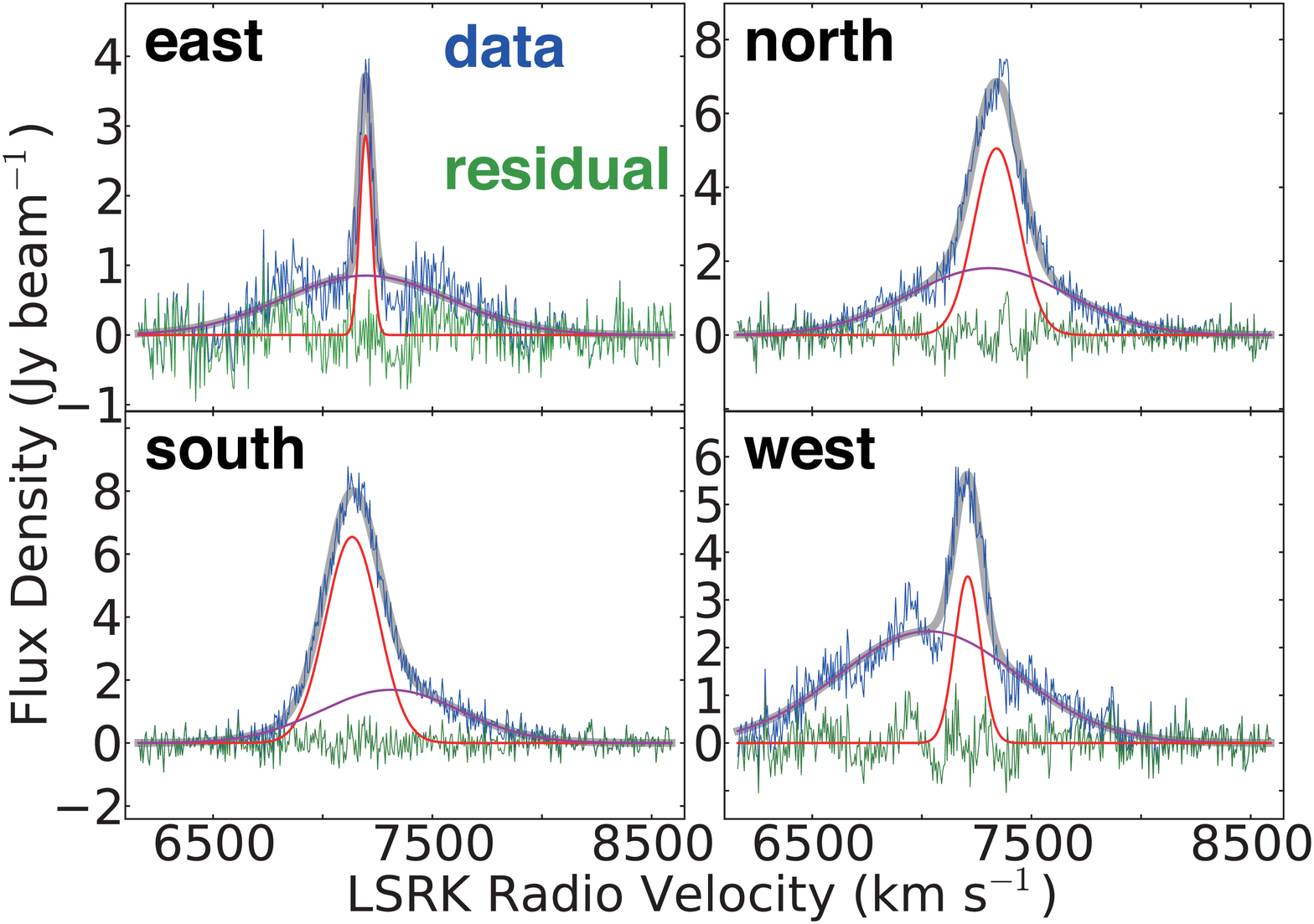}
   \end{center}
    \caption{CO~(2--1) spectrum (blue), residual (green), and two-component Gaussian model (red = narrow, purple = broad, and black = narrow + broad) within 2\farcs0 aperture at the four highest velocity dispersion positions of the CO~(2--1) map.  These four positions are shown in Figure~\ref{fig_3}c.}
    \label{fig_2}
\end{figure}

\section{Results}
The CO~(2--1) integrated intensity contours overlaid on a velocity field map are shown in Figure~\ref{fig_1}a.
It shows a bright nuclear concentration ($\sim$700~pc, a rough size of the contour which has 50\% of the peak value by eye) and extended filamentary structures (length $\sim$ 11~kpc), which coincide with the dust lanes in the optical image \citep{Yoshida16}.
Both features match the CO~(1--0) emission as reported by \citet{Feruglio13a}.
The extended CO emission is compact compared to the H$\alpha$ distribution ($\sim$90~kpc) and the remnant of the main stellar disk ($\sim$20~kpc) \citep{Yoshida16}.
There are extremely broad, asymmetric velocity wings, ranging from $\sim$6200 to $\sim$8200 km s$^{-1}$ in the radio LSRK velocity frame (full width at zero intensity (FWZI) $\sim$ 2000 km s$^{-1}$), within the central 4\arcsec aperture (Figure~\ref{fig_1}a and \ref{fig_1}b).
This linewidth is broader than that previously reported for CO lines \citep[e.g.,][]{Feruglio13a}, because of the higher sensitivity and wider bandwidth of our CO data.
The observed total integrated intensity inside the 3$\sigma$ contour is 1283 $\pm$ 64~Jy km s$^{-1}$, which is 86 $\pm$ 15~\% of the single-dish value \citep{Papadopoulos14}.
We also detected CN~(2$_{5/2}$--1$_{3/2}$) and CS~(3--2) lines, which will be presented in a future paper.

We implemented one and two Gaussian fits to disentangle outflow signatures from the chaotic velocity field of NGC~6240 (Figure~\ref{fig_1}a).
A hexagonal Nyquist sampling with 2\farcs0 aperture (step = 1\farcs0) was performed on the CO~(2--1) cube to decrease the number of spectra (from 500 $\times$ 500 pixels to 40 $\times$ 40 apertures) for the fit.
We fit the data as follows: (1) perform a one-component fit for spectra with a peak S/N $>$ 5 and then (2) if the residual of the one-component fit has a peak S/N $>$ 5, perform a two-component fit to the data (not the residual).
Thus, the resultant best-fit model could have one or two Gaussian component(s) at each aperture.
For the one-component fit, the three initial parameters (i.e., peak, velocity, and FWHM) were chosen from the peak flux, peak velocity, and the number of channels, which satisfy peak S/N $>$ 5.
For the two-component fit, initial parameters of the first Gaussian are obtained from the one-component fit, although those of the second Gaussian were chosen from the residual of the one-component fit following the procedure used for estimating the initial guess of the one-component fit.
Examples of multiple Gaussian fits that require two Gaussian components are shown in Figure~\ref{fig_2}.
The peaks seen in the residual of the two-component fit are statistically insignificant (peak S/N $<$ 5).
Thus, we did not fit any third Gaussian component.

Figure~\ref{fig_3}a shows the histogram of the CO~(2--1) linewidth in FWHM from every Gaussian component from our fitting procedure.
The small linewidth components ($\leq$240 km s$^{-1}$) are mainly found along the optical dust lanes, which is similar to that of giant molecular associations found in the nearby merging galaxy the Antennae \citep[$\leq$210 km s$^{-1}$;][]{Espada12,Ueda12}.
On the other hand, the large linewidth components reach unusually large values of $\sim$1000 km s$^{-1}$.
We performed a Gaussian fit to this plot.
The best-fit Gaussian (green line) shows a peak at 49~km s$^{-1}$ and the dispersion of 160~km s$^{-1}$ (i.e., $+$1$\sigma$ $\simeq$ 210 km s$^{-1}$).
There are 12\% outliers larger than the $+$3$\sigma$ limit (at FWHM $\sim$ 530~km s$^{-1}$) of the Gaussian population, and thus we simply define data whose linewidths are larger than 500~km s$^{-1}$ as ``broad components".
The broad components account for $\lesssim$10\% of the total integrated intensity.
The linewidth image of the broad components is shown in Figure~\ref{fig_3}b and \ref{fig_3}c.
The image area of Figure~\ref{fig_3}b roughly corresponds to that of Figure~\ref{fig_1}a.

\section{Discussion}
The origin of the CO velocity wings in the central region of NGC~6240 has been discussed for years, and recent studies based on interferometric observations suggested the presence of massive molecular outflow(s) (see Section~\ref{S_1}).
However, the spatial and velocity distributions of the molecular outflows are unclear yet mainly because of the complicated CO velocity structures and the instrumental limits.

\begin{figure*}
	\includegraphics[width=2.0\columnwidth]{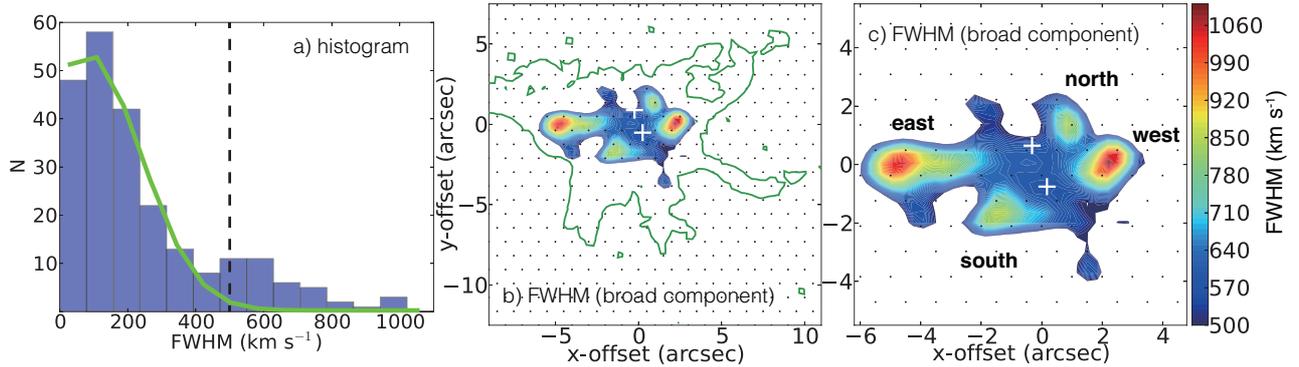}
    \caption{(a) Distribution of the CO~(2--1) linewidth in FWHM (km s$^{-1}$) from every Gaussian component fit.  The green line shows the best fit Gaussian to the histogram, showing that there are outliers higher than 500 km s$^{-1}$ (i.e., broad component). The black dashed line marks where FWHM = 500 km s$^{-1}$.
    (b) CO~(2--1) linewidth map of the broad component in FHWM (km s$^{-1}$). The green contour shows the outline of the H$\alpha$ (``butterfly nebula"). Black dots show the center position of each aperture.
    (c) Zoomed-in CO~(2--1) linewidth map of the broad components with the color bar showing the linewidth range.}
    \label{fig_3}
\end{figure*}

\begin{figure}
	\begin{center}
	\includegraphics[width=0.9\columnwidth]{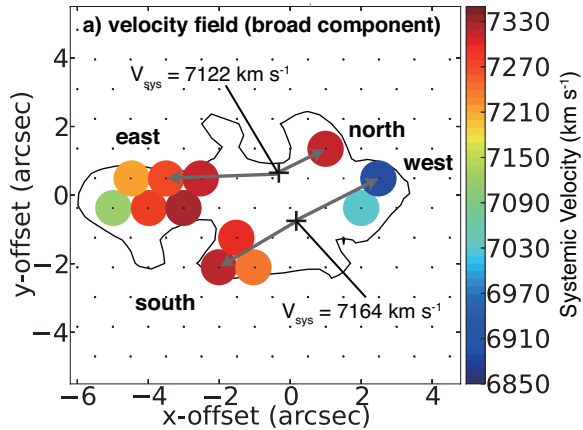}
   \end{center}
    \caption{Peak velocity map of the broad component.  The velocity field in color scale ranges from 6850 to 7350 km s$^{-1}$.  Data points around the four peaks in Figure~\ref{fig_3}c are shown.}
    \label{fig_4}
\end{figure}

The broad component is located around the central 3.4~kpc, although large linewidth regions are not located at the nuclear positions.
The linewidth of the broad components peak at four positions around the nuclei, whose spectra are shown in Figure~\ref{fig_2}.
The broad components show a clearly different spatial distribution with respect to the dense gas tracer, HCN~(4--3), which likely traces nuclear rotating disks \citep{Scoville15}.
The four peak positions of the broad components roughly coincide with peaks of H$\alpha$ \citep{Gerssen04}, near-IR H$_{2}$ \citep{Max05}, and 0.5-1.5 keV X-ray \citep{Komossa03}, indicating that the CO broad components arise from a multi-phase ISM.
This is also supported by the maximum velocity of the CO~(2--1) emission (FWZI/2 $\sim$ 1000 km s$^{-1}$; Figure~\ref{fig_1}b), which is consistent with the maximum velocity of a warm molecular outflow traced by blueshifted OH absorption \citep[$\sim$1200 km s$^{-1}$;][]{Veilleux13} and H$\alpha$ emission \citep[$\geq$1000 km s$^{-1}$;][]{Heckman90}.
As seen in other LIRGs (e.g., NGC~1068; \citealt{Garcia-Burillo14}, NGC~3256; \citealt{Sakamoto14}, and NGC~1614; \citealt{Saito17}), molecular outflows are often spatially extended (a few kpc) and concomitant with (or entrained by) ionized gas outflow.
Considering all evidences described above and some similarities with other LIRGs, we suggest that the four peaks of the broad CO~(2--1) components trace molecular outflows in NGC~6240.

We derive physical parameters of the molecular outflows found in NGC~6240.
The spatially-resolved map (Figure~\ref{fig_3}c) allows us to estimate the mass and projected distance of the linewidth peaks from the launching points.
We assume the launching points are the double nuclei, that is, the northern and eastern (southern and western) components of the molecular outflow come from the northern (southern) nucleus (see gray arrows in Figure~\ref{fig_4}).
This spatial configuration is supported by radio continuum observations with very long baseline interferometers \citep{Gallimore04,Hagiwara11}, showing that the northern nucleus has an east-west bipolar structure.
The direction of those bipolar structures are similar to our assumed morphology of the molecular outflows.
We also assume that the inclination ($\alpha$), CO~(2--1)/CO~(1--0) line intensity ratio, and CO~(1--0) luminosity to H$_2$ mass conversion factor of the molecular outflows in NGC~6240 are 45\degr, unity, and 0.8 $M_{\odot}$ (K km s$^{-1}$ pc$^{2}$)$^{-1}$, respectively, for the sake of simplicity.

To estimate the mass outflow rate ($\dot{M}_{\rm H_2, out}$), we employ the expression,

\begin{equation}
\dot{M}_{\rm H_2, out} = \frac{3v_{\rm out, proj}M_{\rm H_2, out}}{R_{\rm out, proj}}\tan{\alpha},
\end{equation}
where $v_{\rm out, proj}$ is the projected velocity of the outflow, $M_{\rm H_2, out}$ is the gas mass of the outflow, $R_{\rm out, proj}$ is the distance from the launching point to the outflowing gas, and $\alpha$ is the inclination of the outflow.
This equation assumes a uniformly filled cone geometry \citep{Maiolino12}.
All derived parameters related to $\dot{M}_{\rm H_2, out}$ are listed in Table~\ref{table_1}.
$v_{\rm out, proj}$ of the northern and eastern (southern and western) outflows are derived by using the projected velocity of the outflowing gas (Figure~\ref{fig_4}) and the systemic velocity of the northern (southern) nucleus of 7122 (7164) km s$^{-1}$ in the radio LSRK velocity frame \citep{Hagiwara10}.

The derived mass of each broad component is $\sim$10$^{7.9-8.2}$~$M_{\odot}$ ($\sim$10$^{8.7}$~$M_{\odot}$ in total, which is consistent with the value derived from unresolved CO data; \citealt{Cicone14}).
Since the ionized gas mass of the butterfly nebula is $<$1.4 $\times$ 10$^8$~$M_{\odot}$ \citep{Yoshida16}, $>$70 \% of the total gas mass in the outflow consists of molecules.
The total $\dot{M}_{\rm H_2, out}$ of $\sim$230~$M_{\odot}$ yr$^{-1}$ is 3.5 times lower than that estimated by \citet{Cicone14}.
This is due to the difference of the assumed geometry.
\citet{Cicone14} assumed a spherical geometry because their data did not spatially resolve the broad CO component, and we assumed a conical geometry with a certain inclination.
Using the nuclear SFR of $\sim$61~$M_{\odot}$ yr$^{-1}$ derived from radio-to-FIR SED fitting \citep{Yun02}, the mass loading factor ($\dot{M}_{\rm H_2, out}$ divided by SFR) is $\sim$4, indicating that either the AGN, or the starburst, or both are the main drivers of the outflows.
We also estimated the age (= $R_{\rm out, proj}$/$v_{\rm out, proj}$ assuming the inclination of 45\degr) of each broad component, and found that it ranges from 5 to 27 Myr.
This is consistent with the age of the central H$\alpha$ structures from the butterfly nebula (8.4~Myr) to the ``hourglass" (24~Myr) \citep{Yoshida16}, indicating that, again, the H$\alpha$ and CO~(2--1) outflows are colocated.

\begin{table}
	\centering
	\caption{NGC~6240 molecular outflow properties.}
	\label{table_1}
	\begin{tabular}{lccccr} 
		\hline
		Property &east &south &west &north &total \\
		\hline
      $R_{\rm out, proj}$ (pc) &1960 &1240 &1370 &830 &... \\
      $v_{\rm out, proj}$ (km s$^{-1}$) &$-$70 &+150 &$-$270 &+180 &... \\
      age (Myr) &27 &8 &5 &5 &... \\
		$\log M_{\rm H_2, out}$ ($M_{\odot}$) &7.9 &8.1 &8.2 &8.2 &8.7 \\
      $\dot{M}_{\rm H_2, out}$ ($M_{\odot}$ yr$^{-1}$) &8 &43 &86 &94 &231 \\
		\hline
	\end{tabular}
\end{table}

Note that, although the spatial characteristics of the broad CO components in NGC~6240 could also be explained by inflow motions \citep[e.g.,][]{Gaspari17} rather than outflow, we favor the latter because the absorption profile of the OH doublet toward the central 9\arcsec\: shows that a high velocity component faster than 1000~km~s$^{-1}$ is only detected on the blueshifted side \citep{Veilleux13}.

\section{Conclusion}
We present the ALMA high-resolution observations of CO~(2--1) line toward the prototypical ULIRG NGC~6240, which is known to have broad CO profiles around the center, although no one has succeeded in extracting their spatial distributions.
Our high sensitivity and wide-band ALMA data revealed the presence of extremely broad CO wings (FWZI $\sim$ 2000~km s$^{-1}$), which are as fast as the OH and H$\alpha$ outflows.
We performed multiple Gaussian fitting for the CO datacube to visualize the morphology of the wings.
We found that for the first time the broad component presents four peaks 1-2 kpc away from the double nuclei.
We also found a spatial connection between nuclear bipolar structures in radio, CO, and the H$\alpha$ and X-ray emission, and thus we suggest that the CO wings are associated with twin east-west bipolar molecular outflows launched from the nuclei.

\section*{Acknowledgements}
The authors thanks an anonymous referee for comments that improved the contents of this paper.
T.S. thanks T. H. Saitoh and R. Maiolino for useful discussion.
T.S. and the other authors thank the ALMA staff for their kind support.
This paper makes use of the following ALMA data: ADS/JAO.ALMA\#2015.1.00003.S.
ALMA is a partnership of ESO (representing its member states), NSF (USA) and NINS (Japan), together with NRC (Canada), NSC and ASIAA (Taiwan), and KASI (Republic of Korea), in cooperation with the Republic of Chile. The Joint ALMA Observatory is operated by ESO, AUI/NRAO and NAOJ.



\bsp	
\label{lastpage}
\end{document}